\documentclass[12pt]{iopart}
\usepackage{iopams}  
\usepackage{graphicx}

\begin{document}

\eqnobysec
\newtheorem{proposition}{Proposition}
\newtheorem{theorem}{Theorem}


\newcommand\myu{u}

\newcommand\mynum[2]{\ifcase#1 ?\or #2 \or \bar{#2} \else ? \fi}

\newcommand{\myindex}[1]{\if ?#1? \else _{#1}\fi}

\newcommand\mybra[3]{
  \langle\, 
  \ifcase#1 ?\or 
    #2\myindex{#3} \or 
    \bar{#2}\myindex{#3} 
  \else ? \fi \,|}

\newcommand\myket[3]{
  |\, \ifcase#1 ?\or 
    #2\myindex{#3} \or 
    \bar{#2}\myindex{#3} 
  \else ? \fi \,\rangle}

\newcommand\mymat[3]{
  \ifcase#1 ?\or 
    \mathsf{#2}\myindex{#3} \or 
    \bar{\mathsf{#2}}\myindex{#3} \else?\fi}

\newcommand\myshift[1]{\mathbb{T}_{#1}} 

\newcommand\mydiff[1]{\mathbb{D}_{#1}} 

\newcommand\myShifted[3][0]{\ifcase #1
   \mathbb{T}_{#2}{#3} \or                     
   \left(\mathbb{T}_{#2}{#3}\right) \else      
   #3[#2] 
   \fi} 
\newcommand{\myset}[1]{\mathrm{#1}}

\newcommand{\myzero}[3]{
  \if ?#3? \mathfrak{#1}_{#2}\, 
  \else \myShifted{#3}{\mathfrak{#1}_{#2}} \fi}


\title{Solitons of the (2+2)-dimensional Toda lattice.}
\author{V.E. Vekslerchik}
\address{
  Usikov Institute for Radiophysics and Electronics \\
  12, Proskura st., Kharkov, 61085, Ukraine 
}
\ead{vekslerchik@yahoo.com}
\submitto{\JPA}

\begin{abstract}
We use the generalized Cauchy matrix approach to derive the 
N-soliton solutions for the $(2+2)$-dimensional Toda lattice. 
\end{abstract}

\section{Introduction.}

In this note, we discuss the soliton solutions for the $(2+2)$-dimensional 
Toda lattice.

The Toda model, one of the best-studied integrable systems 
which has applications in various fields of mathematics and physics, 
was introduced in \cite{M67} as a nonlinear chain, i.e. 
a $(1+1)$-dimensional system with one continuous (time) 
and one discrete (index) variables. 
One of the questions that has attracted much interest during its fifty-year 
history is to find generalizations of this model, first of all, to describe 
problems in high dimensions (dynamics of two- and three-dimensional lattices, 
coupled fields in two- or three-dimensional spaces, etc).
There are several known three-dimensional extensions of the Toda model: 
the $(2+1)$-dimensional (2 continuous + 1 discrete variable) chains 
\cite{M79,DJM82,UT83a,UT83b,UT84,HIK88}, 
the $(1+2)$-dimensional (1 continuous + 2 discrete variables) lattices  
\cite{DJM82,S02,CC07,WY07}, 
completely discrete  $3$-dimensional lattice \cite{HIK88}. 
However, the attempts to proceed to higher dimensions demonstrate that almost 
all methods of generalization which preserve the integrability 
(with, probably, the only exception \cite{ADMV16}) lead either to 
the models with extra fields \cite{RS97,SND04} (the `matter type fields', in 
the terminology of \cite{RS97}) or to the equations that lose some typical 
features of the integrable systems, like trilinear equations studied in 
\cite{MS90,HS91,MS91}.

The equations we study in this paper can be written as 
\begin{equation}
  \frac{\partial^{2} \myu }{\partial x \partial y} 
  = 
  e^{\Delta_{1} \myu} - e^{\Delta_{2} \myu} 
\label{eq:Toda22}
\end{equation}
where $\myu$ is a function of two continuous ($x$ and $y$) and two discrete 
($n_{1}$ and $n_{2}$) variables, 
\begin{equation}
  \myu = \myu(x,y,n_1,n_2), 
\end{equation}
and $\Delta_{1}$ and $\Delta_{2}$ are the one-dimensional second-order 
difference (discrete Laplace) operators, 
\begin{equation}
  \begin{array}{l} 
  \Delta_{1} \myu(n_1,n_2) 
  = 
  \myu(n_1+1,n_2) - 2\myu(n_1,n_2) + \myu(n_1-1,n_2), 
  \\ 
  \Delta_{2} \myu(n_1,n_2) 
  = 
  \myu(n_1,n_2+1) - 2\myu(n_1,n_2) + \myu(n_1,n_2-1). 
  \end{array}
\end{equation}

Here, we do not discuss the questions related to the integrability of 
\eref{eq:Toda22}, such as the zero-curvature (or Lax) representation, 
the conservation laws, the Hamiltonian structure etc. 
We restrict ourselves, like in the cited above works 
\cite{HIK88,MS90,MS91}, to the problem of finding some class of particular 
solutions. We formulate an \textit{ansatz} and use it to derive the 
N-soliton solutions for \eref{eq:Toda22}.

\section{Soliton matrices. \label{sec:matrices}}

In what follows, we use the soliton matrices that were studied in \cite{V15}. 
We define the soliton tau-functions as the determinants 
\begin{equation}
  \tau  
  = 
  \det\left| \mathsf{1} + \mymat1{A}{} \mymat2{A}{} \right| 
\label{def:tau}
\end{equation}
of the matrices defined by 
\begin{equation}
  \begin{array}{lcl}
  \mymat2{M}{} \mymat1{A}{} - \mymat1{A}{} \mymat1{M}{} 
  & = &
  \myket1{\alpha}{}\mybra1{a}{}, 
  \\[1mm] 
  \mymat1{M}{} \mymat2{A}{} - \mymat2{A}{} \mymat2{M}{} 
  & = &
  \myket2{\alpha}{}\mybra2{a}{}
  \end{array}
\label{def:A}
\end{equation}
where $\mymat1{M}{}$ and $\mymat2{M}{}$ are constant $N \times N$ diagonal 
matrices, 
$\myket1{\alpha}{}$ and $\myket2{\alpha}{}$ are constant $N$-columns 
while  
$\mybra1{a}{}$ and $\mybra2{a}{}$ are $N$-component rows that depend on the 
coordinates describing the model. 
It should be noted that the overbar \emph{does not} indicate the complex 
conjugation: we consider all matrices, rows and columns to be real.

We define shifts $\myshift\zeta$ as 
\begin{equation}
  \begin{array}{l} 
  \myshift\zeta \mybra1{a}{} = \mybra1{a}{} \, \mymat1{H}{\zeta}, 
  \\[2mm] 
  \myshift\zeta \mybra2{a}{} = \mybra2{a}{} \, \mymat2{H}{\zeta} 
  \end{array}
\end{equation}
with 
\begin{equation}
  \begin{array}{l} 
  \mymat1{H}{\zeta} 
  = 
  \left( \mymat1{M}{} - \zeta \right) 
  \left( \mymat1{M}{} + \zeta \right)^{-1}, 
  \\[2mm] 
  \mymat2{H}{\zeta} 
  = 
  \left( \mymat2{M}{} + \zeta \right) 
  \left( \mymat2{M}{} - \zeta \right)^{-1} 
  \end{array}
\label{def:H}
\end{equation}
(we do not indicate the unit matrix explicitly and write 
$\mymat1{M}{} - \zeta$ instead of $\mymat1{M}{} - \zeta\mathsf{1}$, etc) 
which determines the action of the shifts on the matrices $\mymat1{A}{}$ 
and $\mymat2{A}{}$, 
\begin{equation}
  \begin{array}{l} 
  \myshift\zeta \mymat1{A}{} = \mymat1{A}{} \, \mymat1{H}{\zeta}, 
  \\[2mm] 
  \myshift\zeta \mymat2{A}{} = \mymat2{A}{} \, \mymat2{H}{\zeta} 
  \end{array}
\label{eq:shA}
\end{equation}
and, hence, the tau-functions $\tau$. 

Note that the shifts used in this work are different from the ones used in 
\cite{V15}. As the result, the tau-functions \eref{def:tau} satisfy different 
set of equations. What is important is that in the case of \eref{eq:shA} we 
do not have the simple three-point equations that were, in some sense, 
the base of most of the results presented in \cite{V15}. 
Thus, the calculations of this work are somewhat more cumbersome, 
with more emphasis upon the algebraic properties of the matrices \eref{def:A}.

In what follows, we study the results of the combined action of several 
shifts, which we denote as 
\begin{equation}
  \myshift{\xi\eta} = \myshift{\xi}\myshift{\eta}, 
  \quad
  \myshift{\xi\eta\zeta} = \myshift{\xi}\myshift{\eta}\myshift{\zeta}, 
  \quad
  ... 
\end{equation}
or, by means of the set notation, as 
\begin{equation}
  \myset{X} = \left\{ \xi_{1}, \, ... \, , \xi_{N} \right\}, 
  \qquad
  \myshift{\myset{X}}  
  = 
  \prod_{ n =1 }^{N} \myshift{\xi_{n}}. 
\end{equation}

After some simple algebra, one can present the shifted tau-function as 
\begin{equation}
  \myShifted{\xi}{\tau} = E_{\xi\xi} \; \tau 
\label{eq:dtau1}  
\end{equation}
where quantities $E_{\xi\eta}$ are defined by 
\begin{equation}
  E_{\xi\eta}  
  =  
  1 
  + 
  (\xi+\eta) \mybra1{b}{\xi} \mymat1{G}{}\mymat1{A}{} \myket2{\beta}{\eta} 
\end{equation}
with
\begin{equation}
  \mymat1{G}{} = \left( \mathsf{1} + \mymat1{A}{} \mymat2{A}{} \right)^{-1} 
\label{def:G}
\end{equation}
and 
\begin{equation}
  \mybra1{b}{\xi} = \mybra2{a}{} 
  \left( \mymat2{M}{} - \mynum1{\xi} \right)^{-1}, 
\qquad
  \myket2{\beta}{\eta} 
  = 
  \left( \mymat1{M}{} + \mynum1{\eta} \right)^{-1} 
  \myket2{\alpha}{} 
\label{def:ketb2}
\end{equation}
(see \ref{app:Miwa}).

Equation \eref{eq:dtau1} can be generalized to 
\begin{equation}
  \frac{ \myShifted{X}\tau }{ \tau } 
  = 
  \frac{ \mathcal{D}_{X} }{ \mathcal{C}_{X} } 
\label{eq:dtauX}  
\end{equation}
where $\mathcal{D}_{X}$ and $\mathcal{C}_{X}$ are the determinants given by 
\begin{equation}
  \mathcal{D}_{X} = 
  \det\left| \; 
  \frac{ E_{\xi\eta} }{ \xi + \eta }  \; 
  \right|_{\xi,\eta \in X}, 
\hspace{10mm}
  \mathcal{C}_{X} 
  = \det\left| \; \frac{1}{ \xi + \eta } \; \right|_{\xi,\eta \in X} 
\end{equation}
(see \ref{app:Miwa}).

Equation \eref{eq:dtauX} can be used to derive various identities for 
$\myShifted{\myset{X}}{\tau}$ with different $\myset{X}$. 
The ones that we need in this paper can be formulated as the following two 
propositions.

\begin{proposition} \label{prop:Miwa}
The tau-functions \eref{def:tau} satisfy the Miwa 
(discrete BKP) equation 
\begin{equation}
\fl\qquad
  \tau \myShifted{\alpha\beta\gamma}{\tau} 
  = 
    \Gamma_{\alpha,\beta\gamma}
    \myShifted[1]{\alpha}{\tau}
    \myShifted[1]{\beta\gamma}{\tau} 
  + \Gamma_{\beta,\alpha\gamma}
    \myShifted[1]{\beta}{\tau} 
    \myShifted[1]{\alpha\gamma}{\tau}
  + \Gamma_{\gamma,\alpha\beta}
    \myShifted[1]{\gamma}{\tau} 
    \myShifted[1]{\alpha\beta}{\tau}.
\label{eq:Miwa}
\end{equation}
with constants $\Gamma$ defined by  
\begin{equation}
\label{def:Gamma1}
  \Gamma_{\xi,\myset{Y}} 
  =
  \prod_{\eta\in\myset{Y}} 
  \frac{ \xi + \eta }{ \xi - \eta }.
\end{equation}

\end{proposition} 
%
We outline a proof of this statement in \ref{app:Miwa}.

\begin{proposition} \label{prop:MiwaX}
Each solution of equation \eref{eq:Miwa} with \eref{def:Gamma1}
delivers a solution for the second equation of the discrete BKP hierarchy 
\begin{equation}
\fl\qquad
  \tau \myShifted{\alpha\beta\gamma\delta}{\tau} 
  = 
    \Gamma_{\alpha\delta,\beta\gamma}
    \myShifted[1]{\alpha\delta}{\tau}
    \myShifted[1]{\beta\gamma}{\tau} 
  + \Gamma_{\beta\delta,\alpha\gamma}
    \myShifted[1]{\alpha\gamma}{\tau}
    \myShifted[1]{\beta\delta}{\tau} 
  + \Gamma_{\gamma\delta,\alpha\beta}
    \myShifted[1]{\alpha\beta}{\tau}
    \myShifted[1]{\gamma\delta}{\tau} 
\label{eq:MiwaX}
\end{equation}
with constants $\Gamma$ defined by  
\begin{equation}
\label{def:Gamma}
  \Gamma_{\myset{X},\myset{Y}} 
  =
  \prod_{\xi\in\myset{X}} 
  \prod_{\eta\in\myset{Y}} 
  \frac{ \xi + \eta }{ \xi - \eta }. 
\end{equation}

\end{proposition} 
%
We prove this statement in \ref{app:MiwaX}.

Finally, introducing the operators 
\begin{equation}
  \mydiff{\zeta} 
  = 
  \lim_{\alpha\to\zeta}
  \frac{ 1 }{ \varepsilon_{\alpha,\zeta} }
  \left( 
    \myshift{\alpha}\myshift{\zeta}^{-1} - 1 
  \right) 
\label{def:D}
\end{equation}
with 
\begin{equation}
  \varepsilon_{\xi,\eta} 
  = \frac{ \xi - \eta }{ \xi + \eta }
\end{equation}
(note that $\varepsilon_{\alpha,\zeta} \to 0$ as $\alpha\to\zeta$) 
one can derive from equation \eref{eq:MiwaX} and definition \eref{def:Gamma} 
the following result:

\begin{proposition} \label{prop:Toda}
The tau-functions \eref{def:tau} satisfy the bilinear equation 
\begin{eqnarray}
  \tau \left(\mydiff{\lambda}\mydiff{\mu} \tau \right)
  - 
  \left(\mydiff{\lambda} \tau \right)
  \left(\mydiff{\mu} \tau \right)
  & = & 
  \Gamma_{\lambda,\mu}^{2}
      \myShifted[1]{\bar{\lambda}\mu}{\tau}
      \myShifted[1]{\lambda\bar{\mu}}{\tau}
  - 
  \Gamma_{\lambda,\mu}^{-2}
      \myShifted[1]{\bar{\lambda}\bar{\mu}}{\tau}
      \myShifted[1]{\lambda\mu}{\tau} 
\nonumber\\[2mm]&&
  - 
  \left( 
  \Gamma_{\lambda,\mu}^{2} 
  - \Gamma_{\lambda,\mu}^{-2} 
  \right) 
  \tau^{2}
\label{eq:Toda}
\end{eqnarray}
with $\myshift{\bar\zeta}=\myshift{\zeta}^{-1}$. 

\end{proposition} 

This result gives us a possibility to obtain a family of particular solutions 
for \eref{eq:Toda22} by some simple algebraic calculations.

\section{N-soliton  solutions. \label{sec:solitons}}

Let us compare equation \eref{eq:Toda} with fixed $\lambda$ and $\mu$, 
\begin{equation}
  \lambda, \mu = \mbox{constant} 
\end{equation}
(which we consider as two parameters of our solution), 
and the bilinear form of \eref{eq:Toda22}, which can be obtained by the 
substitution $\myu = \ln\omega$, 
\begin{eqnarray}
&&
  \omega\left(n_{1},n_{2}\right) 
  \frac{\partial^{2} \omega\left(n_{1},n_{2}\right) }{\partial{x}\,\partial{y} }
  -
  \frac{\partial \omega\left(n_{1},n_{2}\right) }{\partial{x} }
  \frac{\partial \omega\left(n_{1},n_{2}\right) }{\partial{y} }
\nonumber\\&& \qquad
  = 
  \omega\left(n_{1}-1,n_{2}\right)\omega\left(n_{1}+1,n_{2}\right) 
  - 
  \omega\left(n_{1},n_{2}-1\right)\omega\left(n_{1},n_{2}+1\right), 
\label{eq:Toda22bi}
\end{eqnarray}
where the dependence of $\omega$ on $x$ and $y$ is not indicated explicitly, 
and see how one can modify $\tau$, defined in section \ref{sec:matrices}, to 
transform it into a solution for \eref{eq:Toda22bi}. 

First, it is easy to note that the first two terms in the right-hand side of 
\eref{eq:Toda} coincide, up to the constants, with the right-hand side of 
\eref{eq:Toda22bi} provided we introduce the dependence on $n_{1}$ and $n_{2}$ 
in such a way that the translations $n_{1} \to n_{1}+1$ and $n_{2} \to n_{2}+1$ 
lead to the same result as application of the shifts 
$\myshift{\lambda}\myshift{\mu}^{-1}$ and $\myshift{\lambda}\myshift{\mu}$. 
This is easy to achieve by introducing the $(n_{1},n_{2})$-dependence of the 
rows $\mybra1{a}{}$ and $\mybra2{a}{}$ 
(and hence of the matrices $\mymat1{A}{}$ and $\mymat2{A}{}$) as follows: 
\begin{equation}
\begin{array}{l} 
  \mybra1{a}{} 
  = 
  \langle\, a(n_{1},n_{2}) \,|
  = 
  \langle\, c \,|
  \mymat1{H}{}_{\lambda}^{n_{1}+n_{2}} 
  \mymat1{H}{}_{\mu}^{n_{2}-n_{1}}, 
\\[2mm] 
  \mybra2{a}{} 
  = 
  \langle\, \bar{a}(n_{1},n_{2}) \,|
  = 
  \langle\, \bar{c} \,|
  \mymat2{H}{}_{\lambda}^{n_{1}+n_{2}} 
  \mymat2{H}{}_{\mu}^{n_{2}-n_{1}} 
\end{array}
\end{equation}
with constant $\langle\, c \,|$ and $\langle\, \bar{c} \,|$.

Next, it should be noted that the operators $\mydiff{\zeta}$ defined in 
\eref{def:D} are, in fact, differential operators. Thus, it is possible 
to introduce the $x$- and $y$-dependence 
of the rows $\mybra1{a}{}$ and $\mybra2{a}{}$ 
(and hence of the matrices $\mymat1{A}{}$ and $\mymat2{A}{}$) 
so that the action of $\mydiff{\lambda}$ and $\mydiff{\mu}$ 
defined in terms of the $\myshift{}$-shifts lead to the same results as the 
differentiating with respect to $x$ and $y$. 
Applying $\mydiff{\zeta}$ to $\mybra1{a}{}$ and $\mybra2{a}{}$,
\begin{equation}
  \mydiff{\zeta} \mybra1{a}{} = \mybra1{a}{} \mymat1{L}{\zeta}, 
  \quad
  \mydiff{\zeta}\mybra2{a}{} = \mybra2{a}{} \mymat2{L}{\zeta} 
\end{equation}
with 
\begin{equation}
\label{def:L}
\begin{array}{l}
  \mymat1{L}{\zeta}  
  = 
  \lim\limits_{\alpha\to\zeta}
  \frac{ \alpha + \zeta }{ \alpha - \zeta }
  \left( 
    \mymat1{H}{\alpha}\mymat1{H}{}_{\zeta}^{-1} - \mathsf{1} 
  \right) 
  = 
  \mymat1{H}{\zeta} - \mymat1{H}{}_{\zeta}^{-1}, 
\\[4mm] 
  \mymat2{L}{\zeta}  
  = 
  \lim\limits_{\alpha\to\zeta}
  \frac{ \alpha + \zeta }{ \alpha - \zeta }
  \left( 
    \mymat2{H}{\alpha}\mymat2{H}{}_{\zeta}^{-1} - \mathsf{1} 
  \right) 
  = 
  \mymat2{H}{\zeta} - \mymat2{H}{}_{\zeta}^{-1}
\end{array}
\end{equation}
and taking  
\begin{equation}
\begin{array}{l}
  \langle\, a \,|
  = 
  \langle\, a(x,y,n_{1},n_{2}) \,|
  = 
  \langle\, a(n_{1},n_{2}) \,|
  \exp\left( x \mymat1{L}{\lambda} + y \mymat1{L}{\mu} \right), 
\\[2mm] 
  \langle\, \bar{a} \,|
  = 
  \langle\, \bar{a}(x,y,n_{1},n_{2}) \,|
  = 
  \langle\, \bar{a}(n_{1},n_{2}) \,|
  \exp\left( x \mymat2{L}{\lambda} + y \mymat2{L}{\mu} \right) 
\end{array}
\end{equation}
we ensure 
$\mydiff{\lambda}\mybra1{a}{} = \frac{\partial}{\partial x} \mybra1{a}{}$ 
and
$\mydiff{\mu}\mybra1{a}{} = \frac{\partial}{\partial y} \mybra1{a}{}$ 
with similar result for $\mybra2{a}{}$.

Finally, one has to take into account the factors 
$\Gamma_{\lambda,\mu}^{\pm 2}$ in the first two terms and to `eliminate' the 
last two terms in the right-hand side of \eref{eq:Toda}.
This can be done by introducing a 
function $\varphi$, 
\begin{equation}
  \omega = e^{\varphi} \tau, 
\end{equation}
which satisfies 
\begin{equation}
\begin{array}{l}
  \Delta_{1} \varphi = \phantom{-}  2 \ln\left|\Gamma_{\lambda,\mu}\right|, 
  \\[2mm] 
  \Delta_{2} \varphi = - 2 \ln\left|\Gamma_{\lambda,\mu}\right|, 
  \\[2mm] 
  \frac{\partial^{2}\varphi }{ \partial{x}\partial{y} }
  = 
  \Gamma_{\lambda,\mu}^{2} - \Gamma_{\lambda,\mu}^{-2}. 
\end{array}
\label{eq:phi}
\end{equation}

Now we can formulate the main result of this paper.

\begin{theorem} 
\label{prop:main}
The N-soliton solutions for the $(2+2)$-dimensional Toda lattice 
\eref{eq:Toda22} can be written as 
\begin{equation}
\fl\qquad 
  \myu\left(x,y,n_{1},n_{2}\right) = 
  \varphi\left(x,y,n_{1},n_{2}\right) 
  + 
  \ln\det\left| 
    \mathsf{1} 
    + 
    \mymat1{A}{}\left(x,y,n_{1},n_{2}\right) 
    \mymat2{A}{}\left(x,y,n_{1},n_{2}\right) 
  \right|
\label{solution}
\end{equation}
where 
\begin{equation}
\fl\qquad 
  \varphi\left(x,y,n_{1},n_{2}\right) 
  = 
  8 xy 
  \frac{\lambda\mu\left( \lambda^{2} + \mu^{2} \right)}
       {\left( \lambda^{2} - \mu^{2} \right)^{2}} 
  + 
  \left( n_{1}^{2} - n_{2}^{2} \right) 
  \ln\left| 
    \frac{ \lambda + \mu }{ \lambda - \mu }
  \right|,
\label{rslt:phi}
\end{equation}
the matrices $\mymat1{A}{}(x,y,n_{1},n_{2})$ and 
$\mymat2{A}{}(x,y,n_{1},n_{2})$ are given by 
\begin{equation}
\begin{array}{l} 
  \mymat1{A}{}(x,y,n_{1},n_{2}) 
  = 
  \mymat1{A}{0} 
  \exp\left( x \mymat1{L}{\lambda} + y \mymat1{L}{\mu} \right) 
  \mymat1{H}{}_{\lambda}^{n_{1}+n_{2}} 
  \mymat1{H}{}_{\mu}^{n_{2}-n_{1}}, 
\\[2mm]
  \mymat2{A}{}(x,y,n_{1},n_{2}) 
  = 
  \mymat2{A}{0} 
  \exp\left( x \mymat2{L}{\lambda} + y \mymat2{L}{\mu} \right) 
  \mymat2{H}{}_{\lambda}^{n_{1}+n_{2}} 
  \mymat2{H}{}_{\mu}^{n_{2}-n_{1}} 
\end{array}
\end{equation}
with the matrices 
$\mymat1{H}{\lambda}$, $\mymat2{H}{\lambda}$, 
$\mymat1{L}{\lambda}$ and $\mymat2{L}{\lambda}$ 
defined in \eref{def:H} and \eref{def:L}, 
and the elements of $\mymat1{A}{0}$ and $\mymat2{A}{0}$ given by 
\begin{equation}
  \left(\mymat1{A}{0}\right)_{j,k} = 
    \frac{ \mynum1{c}_{k} }{\mynum2{M}_{j} - \mynum1{M}_{k} }, 
  \qquad j,k=1, ..., N, 
\end{equation}
\begin{equation}
  \left(\mymat2{A}{0}\right)_{j,k} = 
    \frac{ \mynum2{c}_{k} }{\mynum1{M}_{j} - \mynum2{M}_{k} }, 
  \qquad j,k=1, ..., N. 
\end{equation}
Here, 
$\lambda$, $\mu$, 
$\mynum1{M}_{j}$, $\mynum2{M}_{j}$, 
$\mynum1{c}_{j}$, $\mynum2{c}_{j}$ 
($j=1, ..., N$) 
are arbitrary constants that play the role of parameters in \eref{solution}.
\end{theorem} 
Note that we have put all components of the columns 
$\myket1{\alpha}{}$ and $\myket2{\alpha}{}$ equal to the unity, which can be 
done without loss of generality by redefining the constants 
$\mynum1{c}_{k}$ and $\mynum2{c}_{k}$ 
(the components of the rows $\mybra1{c}{}$ and $\mybra2{c}{}$). 
Also, we have written $\varphi$ as the simplest solution of \eref{eq:phi}. 
One can, in principle, `generalize' \eref{rslt:phi} by adding two functions 
of one variable,  $\varphi_{1}(x)$ and $\varphi_{2}(y)$,  
as well as terms proportional to $n_{1}$, $n_{2}$ and $n_{1}n_{2}$, 
which is a manifestation of the trivial symmetries of \eref{eq:Toda22}.

Of $4N+2$ constants mentioned in Theorem \ref{prop:main}, the parameters 
$\lambda$ and $\mu$ play a special role. 
They completely determine the background 
solution $\varphi$ which completely determines the asymptotic behaviour of 
$\myu$. Even if we `forget' about the background and consider 
$\tilde\myu = \myu - \varphi$, then, by changing the values of $\lambda$ and 
$\mu$ with respect to the values of $\mynum1{M}_{j}$ and $\mynum2{M}_{j}$, one 
can control the signs of the elements of the 
$\mymat1{L}{}$-  and $\mymat2{L}{}$-matrices and the moduli (compared with the 
unity) of the elements of the 
$\mymat1{H}{}$-  and $\mymat2{H}{}$-matrices that, in its turn, determines which 
elements of the $\mymat1{A}{}$-  and $\mymat2{A}{}$-matrices 
grow or vanish in which sector of the asymptotic region, i.e. the asymptotic 
properties of $\tilde\myu$.

\subsection{One-soliton case. }

Here, we discuss the simplest soliton solutions. 
First, let us consider the one-soliton case. 
After assuming 
$\mynum1{c}_{1}\mynum2{c}_{1} < 0$ 
(which ensures absence of singularities) 
and imposing a technical restriction 
$
  \left( \mynum1{M}_{1}^{2} - \lambda^{2} \right) 
  \left( \mynum2{M}_{1}^{2} - \lambda^{2} \right) 
  \left( \mynum1{M}_{1}^{2} - \mu^{2} \right) 
  \left( \mynum2{M}_{1}^{2} - \mu^{2} \right) > 0, 
$
(which excludes sign alternating) 
the function $\myu$ can be presented as 
\begin{equation}
  \myu\left(x,y,n_{1},n_{2}\right) = 
  \varphi\left(x,y,n_{1},n_{2}\right) 
  + 
  \ln\det\left| 
    1 
    + 
    e^{2 f\left(x,y,n_{1},n_{2}\right)} 
  \right| 
\end{equation}
where $f$ is a linear, with respect to all its arguments, function,  
\begin{equation}
  f\left(x,y,n_{1},n_{2}\right) 
  = 
  f_{0} + k_{x} x + k_{y} y + \gamma_{1} n_{1} + \gamma_{2} n_{2}, 
\end{equation}
$f_{0}$ is an arbitrary constant while the constants 
$k_{x}$, $k_{y}$, $\gamma_{1}$ and $\gamma_{2}$ 
(which are combinations of 
$\mynum1{M}_{1}$, $\mynum2{M}_{1}$, $\mynum1{c}_{1}$, $\mynum2{c}_{1}$, 
$\lambda$ and $\mu$ that are not presented here explicitly)
satisfy the `dispersion relation' 
\begin{equation}
  k_{x}k_{y} 
  = 
  \Gamma_{\lambda,\mu}^{2} \sinh^{2}\gamma_{1} 
  - 
  \Gamma_{\lambda,\mu}^{-2} \sinh^{2}\gamma_{2}. 
\end{equation}
It is easy to see that, even if we forget about the background $\varphi$, 
the behaviour of the soliton part of the solution, $\myu - \varphi$, is different 
from what is expected of a soliton: 
\begin{equation}
  \myu - \varphi \sim 
  \left\{ 
  \begin{array}{lll}
	e^{ - 2 |f| } & \to 0      & \mbox{as} \; f \to - \infty \\
	2f            & \to \infty & \mbox{as} \; f \to + \infty 
  \end{array}
  \right..
\end{equation}
However, after calculating the second derivatives one arrives at the famed 
$\mbox{sech}$-expression:
\begin{equation}
  \frac{\partial^{2} }{\partial x \partial y} 
  \left( \myu - \varphi \right) 
  = 
  \frac{ k_{x}k_{y} }{ \cosh^{2} f }. 
\label{eq:uxy-one}
\end{equation}
Similarly, one can derive the `standard' soliton formulae for the second-order 
differences:
\begin{equation}
  \begin{array}{l} \displaystyle 
  e^{\Delta_{1} \myu} 
  = 
  \Gamma_{\lambda,\mu}^{2}
  \left[ 1 + \frac{ \sinh^{2}\gamma_{1} }{ \cosh^{2} f } \right], 
  \\[4mm] \displaystyle  
  e^{\Delta_{2} \myu} 
  = 
  \Gamma_{\lambda,\mu}^{-2}
  \left[ 1 + \frac{ \sinh^{2}\gamma_{2} }{ \cosh^{2} f } \right]. 
  \end{array}
\end{equation}

\subsection{Two-soliton case. }

To demonstrate the structure of the two-soliton solutions we calculate 
\eref{solution} for some fixed set of soliton parameters:
$\mynum1{M}_{1}=7.25$, $\mynum2{M}_{1}=9.25$, 
$\mynum1{c}_{1}=1.0$, $\mynum2{c}_{1}=-2.0$,
$\mynum1{M}_{2}=1.75$, $\mynum2{M}_{2}=2.25$, 
$\mynum1{c}_{2}=1.0$, $\mynum2{c}_{2}=-0.5$,
$\lambda = 1$ and $\mu = 10$.
To make the plots more clear we present there not the function $\myu$ itself 
but the second derivative of its soliton part (without the background 
$\varphi$), 
\begin{equation}
  w_{s} = 
  \frac{\partial^{2} }{\partial x \partial y} 
  \left( \myu - \varphi \right), 
\end{equation}
which in the one-soliton case is given by \eref{eq:uxy-one}.

\begin{figure}%
\begin{center}
\includegraphics{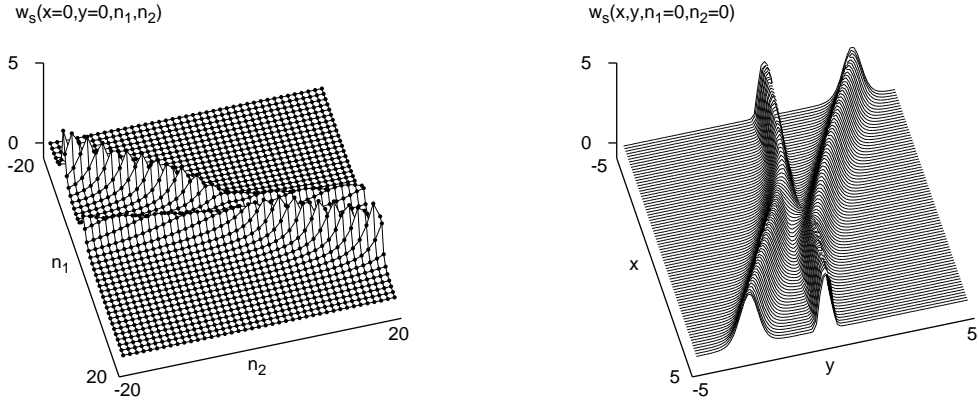}%
\end{center}
\caption{%
The $(n_{1},n_{2})$-dependence (left) and the $(x,y)$-dependence (right) of 
the two-soliton solution.
}%
\label{fig-2}%
\end{figure}

It is easy to see from figure \ref{fig-2} that with continuous coordinates 
being fixed we have typical two-discrete soliton configuration and, 
correspondingly,
with discrete coordinates being fixed we have typical two-continuous soliton 
configuration. In both cases, there is superposition of two solitons in the 
asymptotical regions with typical soliton shifts in the zones of crossings.

\begin{figure}%
\begin{center}
\includegraphics{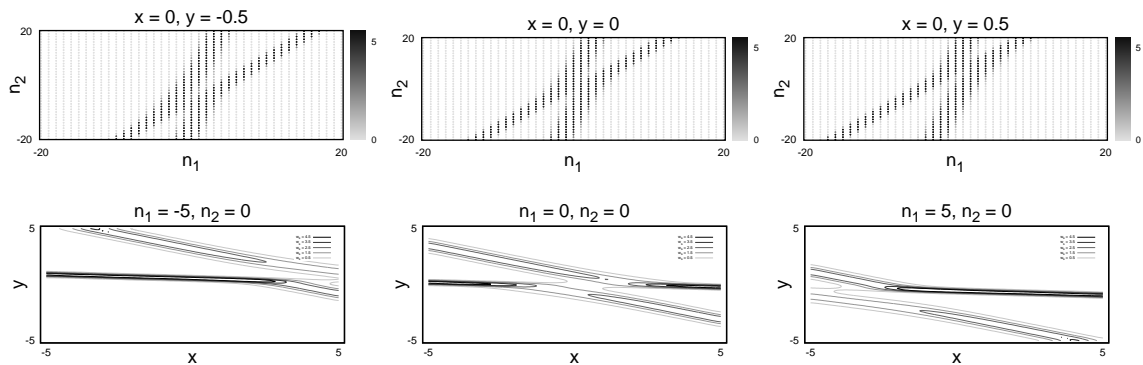}%
\end{center}
\caption{%
The $(n_{1},n_{2})$-distribution for different values of the continuous 
variables (top row) 
and 
the $(x,y)$-distribution for different values of the discrete  
variables (bottom row) 
}%
\label{fig-3}%
\end{figure}

In figure \ref{fig-3} we illustrate the interplay between the discrete and 
continuous coor\-dinates. The most obvious effect is the shift of the 
distributions as a whole. Of course, this effect is not the only one: 
the different dependence of different elements of the matrices 
$\mymat1{A}{}$ and $\mymat2{A}{}$ can noticeably change the value of the 
determinant in \eref{solution} and hence the form of the solution.

\section{Discussion. }

As one can see from the above presentation, we have derived the N-soliton 
solutions using the well-known construction based on matrices \eref{def:A}
which are the so-called `almost-intertwining' matrices, or matrices that 
satisfy the `rank one condition' which is a particular case of the Sylvester 
equation (see \cite{KG01,GK02,GK06,NAH09,DM10a,DM10b,ZZ13,XZZ14}). 

The procedure we have used can be viewed as a version of the 
direct linearization method based on the Cauchy matrices
(see \cite{NQC83,QNCL84}, the book \cite{HJN16} and references therein).
From the viewpoint of the discrete BKP equation 
(see proposition \ref{prop:Miwa}), our generalization consists in using the 
product of the matrices $\mymat1{A}{}$ and $\mymat2{A}{}$ instead of only one 
of them, as in \cite{FN17}, where the authors obtained, 
in the framework of the direct linearizing transform, 
the so-called pfaffian solutions \cite{TH96}. 

The appearance of the pair $\mymat1{A}{}$ and $\mymat2{A}{}$ is a 
characteristic feature of the complex models like the Ablowitz-Ladik (discrete 
nonlinear Schr\"odinger) equations \cite{AL75} (see section 6 of \cite{V15}). 
However, the relationship between the complex Ablowitz-Ladik equations and the 
real Toda model is not new. In some sense, the result of this paper can be 
viewed as an extension of the one of \cite{V95} where the correspondence 
between the $(2+1)$-dimensional Toda lattice and the Ablowitz-Ladik hierarchy 
has been discussed.

Finally, we would like to make a comment on the use of the word `soliton' 
throughout this paper. We have employed the typical soliton ansatz: if we 
expand the determinant in \eref{def:tau}, we arrive at the sum of 
$\exp$-functions which is a version of the Hirota ansatz. Thus, from this 
viewpoint, we have typical soliton tau-functions and, hence, the solutions 
presented in this paper are solitons. On the other hand, 
$\myu\left(x,y,n_{1},n_{2}\right)$ given by \eref{solution} grows in almost 
all directions, while the solitons are usually expected to vanish at the 
infinity or, at least, be bounded (as dark solitons). Nevertheless, in our 
opinion, such solutions (constructed of the soliton tau-functions but 
unbounded) may still be termed as `solitons' without leading to much confusion.

\appendix 

\section{Proof of proposition \ref{prop:Miwa}. \label{app:Miwa}}

Application of \eref{def:A} together with \eref{eq:shA} and \eref{def:H} leads 
to 
\begin{equation}
  \left(  \myshift{\zeta} - 1 \right) \mymat1{A}{}\mymat2{A}{}
  = 
  2\zeta 
  \myket2{\gamma}{\zeta} 
  \mybra1{b}{\zeta} 
\label{eq:shAA}
\end{equation}
which implies
\begin{eqnarray}
  \myShifted{\zeta}{\tau} 
  & = & 
  \det\left| 
    \mathsf{1} 
    + \mymat1{A}{}\mymat2{A}{} 
    + 2 \zeta \myket2{\gamma}{\zeta} \mybra1{\beta}{\zeta} 
  \right|
\\
  & = & 
  \tau 
  \det\left| 
    \mathsf{1} 
    + 2 \zeta \mymat1{G}{} \myket2{\gamma}{\zeta} \mybra1{\beta}{\zeta} 
  \right|
\label{eq:a1}
\end{eqnarray}
where $\mymat1{G}{}$ is defined in \eref{def:G} and 
\begin{equation}
  \myket2{\gamma}{\mu} 
  = \mymat1{A}{} \myket2{\beta}{\mu}. 
\label{def:ketc2}
\end{equation}
Calculating the determinant of the `almost rank-one' matrix in \eref{eq:a1} 
one arrives at \eref{eq:dtau1}. 

Using \eref{eq:shA} and \eref{eq:shAA} we can calculate the `evolution' of 
$\mymat1{G}{}$, $\mybra1{b}{\lambda}$ and $\myket2{\gamma}{\mu}$:
\begin{equation}
  \myshift{\zeta} \mymat1{G}{} 
  = 
  \mymat1{G}{} 
  - 
  2\zeta 
  E_{\zeta\zeta}^{-1} \, 
  \mymat1{G}{} \myket2{\gamma}{\zeta} \mybra1{b}{\zeta} \mymat1{G}{}, 
\end{equation}
\begin{equation}
  (\zeta-\lambda) \; \myshift\zeta \mybra1{b}{\lambda} 
  =  
  2\zeta  \mybra1{b}{\zeta}
  - (\zeta+\lambda) \mybra1{b}{\lambda}, 
\end{equation}
\begin{equation}
  (\zeta-\mu) \; \myshift\zeta \myket2{\gamma}{\mu} 
  = 
  2\zeta \myket2{\gamma}{\zeta}
  - (\zeta+\mu) \myket2{\gamma}{\mu}
\end{equation}
which leads to 
\begin{equation}
\fl\qquad
  (\zeta-\lambda)(\zeta-\mu) \; 
  E_{\zeta\zeta} \; 
  \myshift\zeta E_{\lambda\mu} 
  = 
  (\zeta+\lambda)(\zeta+\mu) \; 
  E_{\zeta\zeta} \; E_{\lambda\mu} 
  - 
  2\zeta (\lambda+\mu) \; 
  E_{\lambda\zeta} \; E_{\zeta\mu} 
\end{equation}
and then 
(after replacing $\lambda,\mu \to \xi$ and $\zeta\to\eta$)
to the two-shift determinant formula 
\begin{equation}
  \frac{ \myShifted{\xi\eta}{\tau} }{ \tau }
  = 
  - 4\xi\eta \, \Gamma_{\xi,\eta}^{2} 
  \left|
  \begin{array}{cc} 
    D_{\xi\xi} & D_{\xi\eta} \\ D_{\eta\xi} & D_{\eta\eta} 
  \end{array}
  \right|,
\qquad
  D_{\xi\eta} = \frac{ E_{\xi\eta} }{ \xi + \eta }. 
\label{eq:dtau2}
\end{equation} 
This relation can be recursively generalized to the $N$-determinants which 
yields the $N$-shift formulae \eref{eq:dtauX}. 
The last fact that we need to prove the proposition \ref{prop:Miwa} is not 
obvious, though rather easy to demonstrate: the functions of two variables 
$E_{\lambda\mu}$ can be factorized into the scalar products of two-vectors, 
each of which depends on only one of them. 
Indeed, by simple algebra one can verify the identity 
\begin{equation}
  E_{\lambda\mu} 
  = 
  \left( \vec{\psi}_{\lambda}, \vec{\phi}_{\mu} \right) 
\end{equation}
with 
\begin{equation}
  \vec{\psi}_{\lambda} 
  = 
  \left(
  \begin{array}{c} 
  1 - \mybra1{a}{} \mymat2{G}{} \mymat2{A}{} \myket1{\beta}{\lambda} 
  \\ 
  \mybra2{a}{} \mymat1{G}{} \myket1{\beta}{\lambda} 
  \end{array}
  \right), 
\qquad
  \vec{\phi}_{\mu} 
  = 
  \left(
  \begin{array}{c} 
  1 - \mybra2{a}{} \mymat1{G}{} \mymat1{A}{} \myket2{\beta}{\mu} 
  \\
  \mybra1{a}{} \mymat2{G}{} \myket2{\beta}{\mu} 
  \end{array}
  \right)  
\end{equation}
where 
\begin{equation}
  \mymat2{G}{} = \left( \mathsf{1} + \mymat2{A}{} \mymat1{A}{} \right)^{-1}, 
\qquad
  \myket1{\beta}{\lambda} 
  = 
  \left( \mymat2{M}{} - \mynum1{\lambda} \right)^{-1}  \myket1{\alpha}{}, 
\end{equation}
while $\mymat1{G}{}$ and $\myket2{\beta}{\mu}$ are defined in \eref{def:G} 
and \eref{def:ketb2}. Since any three two-vectors are linearly dependent, 
it is easy to conclude that 
\begin{equation}
  \det\left| \; 
  E_{\xi_{j}\eta_{k}} \; 
  \right|_{j,k=1,...,N} 
  = 0
  \qquad
  N \ge 3. 
\label{eq:EX}
\end{equation}

Now, to prove the fact that $\tau$ satisfies \eref{eq:Miwa} one has just to 
write \eref{eq:dtauX} with $\myset{X}=\{\xi,\eta,\zeta\}$, 
to substitute 
$E_{\xi\eta}E_{\eta\zeta}E_{\zeta\xi}+E_{\xi\zeta}E_{\zeta\eta}E_{\eta\xi}$ 
from \eref{eq:EX} with 
$\{ \xi_{1},\xi_{2},\xi_{3} \}=\{\xi,\eta,\zeta\}$ 
and to rewrite the remaining combinations $E_{\xi\eta}E_{\eta\xi}$ and 
$E_{\xi\xi}$ using \eref{eq:dtau2} and \eref{eq:dtau1}.

\section{Proof of proposition \ref{prop:MiwaX}. \label{app:MiwaX}}

The proof of proposition \ref{prop:MiwaX} is based on simple algebraic 
calculations. It is straight\-forward to show that the bilinear combination of 
the tau-functions that appears in \eref{eq:MiwaX}, 
\begin{eqnarray}
  \myzero{h}{\alpha\beta\gamma\delta}{}
  & = & 
    \Gamma_{\alpha\delta,\beta\gamma}
    \myShifted[1]{\alpha\delta}{\tau}
    \myShifted[1]{\beta\gamma}{\tau} 
  + \Gamma_{\beta\delta,\alpha\gamma}
    \myShifted[1]{\alpha\gamma}{\tau}
    \myShifted[1]{\beta\delta}{\tau} 
\nonumber\\&&
  + \Gamma_{\gamma\delta,\alpha\beta}
    \myShifted[1]{\alpha\beta}{\tau}
    \myShifted[1]{\gamma\delta}{\tau} 
  - \tau
    \myShifted{\alpha\beta\gamma\delta}{\tau} 
\end{eqnarray}
is a linear combination, 
\begin{eqnarray}
\fl\qquad
    \myShifted[1]{\delta}{\tau}
    \myzero{h}{\alpha\beta\gamma\delta}{}
  & = & 
  + \Gamma_{\alpha,\beta\gamma}
    \myShifted[1]{\alpha\delta}{\tau}
    \myzero{f}{\beta\gamma\delta}{} 
  + \Gamma_{\beta,\alpha\gamma}
    \myShifted[1]{\beta\delta}{\tau}
    \myzero{f}{\alpha\gamma\delta}{} 
  + \Gamma_{\gamma,\alpha\beta}
    \myShifted[1]{\gamma\delta}{\tau}
    \myzero{f}{\alpha\beta\delta}{} 
\nonumber\\&&
  + \tau \myzero{f}{\alpha\beta\gamma}{\delta}, 
\end{eqnarray}
of the bilinear expressions $\myzero{f}{\xi\eta\zeta}{}$, 
\begin{equation}
\fl\qquad
    \myzero{f}{\xi\eta\zeta}{}
  = 
    \Gamma_{\xi,\eta\zeta}
    \myShifted[1]{\xi}{\tau}
    \myShifted[1]{\eta\zeta}{\tau} 
  + \Gamma_{\eta,\xi\zeta}
    \myShifted[1]{\eta}{\tau}
    \myShifted[1]{\xi\zeta}{\tau} 
  + \Gamma_{\zeta,\xi\eta}
    \myShifted[1]{\zeta}{\tau}
    \myShifted[1]{\xi\eta}{\tau} 
  - \tau
    \myShifted{\xi\eta\zeta}{\tau}, 
\end{equation}
that constitute the Miwa equation (see \eref{eq:Miwa}). 
Thus, 
\begin{equation}
  \myzero{f}{\xi\eta\zeta}{} = 0
  \quad
  (\forall \xi,\eta,\zeta) 
  \quad \Rightarrow \quad
  \myzero{h}{\alpha\beta\gamma\delta}{} = 0
\end{equation}
i.e., if $\tau$ is a solution of the \eref{eq:Miwa} 
($\myzero{f}{\xi\eta\zeta}{} = 0$), then 
it is a solution of the \eref{eq:MiwaX} 
($\myzero{h}{\alpha\beta\gamma\delta}{} = 0$).


\end{document}